# On a time dependent cosmological constant

Hristu Culetu[*]

Ovidius University, Dept. of Physics,

B-dul Mamaia 124, 900527 Constanta, Romania

PACS : 98.80. Es

**Abstract**

A time dependent "cosmological constant" $\Lambda(t)$ is conjectured, in terms of the Gaussian curvature of the causal horizon. It is nonvanishing even in Minkowski space because of the lack of informations beyond the light cone. Using the Heisenberg Principle, the corresponding energy of the quantum fluctuations localized on the past or future null horizons is proportional to $\Lambda^{1/2}$.
We compute $\Lambda(t)$ for the (Lorenzian version) of the (conformally flat) Hawking wormhole geometry (written in static spherical Rindler coordinates) and for the de Sitter spacetime.

[*]e-mail : hculetu@yahoo.com

In recent years much attention has been paid to express the cosmological constant in terms of the vacuum energy of some quantum fields [1-7].

F.Mansouri [1] proposed an intimate relation between the vacuum energy and the (cosmological) time dependent $\Lambda(t)$. His main idea is that a high energy experiment modifies the local structure of spacetime for a short period of time, when the Minkowski space becomes dS or AdS and the energy scale is associated with their radius of curvature.

M.R.Setare and R.Mansouri [2] calculared the Casimir stress on a spherical shell in the conformally-flat de Sitter background for a massless scalar field with different vacua out- and inside the shell, representing a bubble in the early Universe.

Jack Ng and H. van Dam [3], based on "unimodular gravity", took $\Lambda$ conjugate to the spacetime volume $\Omega$ of the Universe. Their $\Lambda$ fluctuates about zero with the magnitude $\Omega^{-1/2}$, namely $\Lambda \Omega^{-1/2} \sim 1$ (a Heisenberg type uncertainty principle).

A similar conjecture is expressed by R.G. Vishwakarma [4] who assumed that $\Lambda$ is a stochastic variable arising from quantum fluctuations. It is the rms fluctuation which is observed at the cosmological level.

A.Gregori [5] assumed, as a consequence of String (or M) Theory, that all coordinates are compact and bounded by the horizon of observations (we do not know what happens in regions causally disconnected from our one). His opinion is that the cosmological constant is nothing but the manifestation of the Uncertainty Principle on a cosmological scale. It turns out to be a function of time owing to the expansion of the horizon.

R.R.Caldwell [6] related $\Lambda$ to zero point fluctuations of vacuum fields (Casimir effect). He expressed the idea that, if $\Lambda$ is due to energy then it is susceptible to fluctuations induced by grtavitational forces.

According to M. Ahmed et al. [7], the cosmological constant can fluctuate with a magnitude that diminishes as the Universe grows older. Therefore, they consider $\Lambda$ to be conjugate to the spacetime volume $\Omega$, as for energy and time in Quantum Mechanics. Specifically, we have $\Delta\Lambda \sim \Omega^{-1/2}$ because the fluctuations must be of Poisson type. The authors of [7] identified $\Omega$, which governs the magnitude of fluctuations in time, with the volume of the past light cone.

We conjecture in this letter a "cosmological constant" $\Lambda$ in terms of the Gaussian curvature of the causal horizon ($\Lambda^+$ and $\Lambda^-$ for the future and past null rays, respectively). Our point of view is that $\Lambda$ depends on the spacetime we are living in.

The units will be such that $c = h = G = 1$, excepting special cases.

Let us take an inertial observer who decided to make a measurement in a time $\Delta t$. For causality reasons, he is constrained to extend the measurement on a distance less than $\Delta t$, the distance travelled by light during the experiment. We may say that our observer lives in a spherical box of radius $\Delta t$, the surface of the sphere acting as a "knowledge horizon" due to the lack of informations beyond it, during the experiment. As the horizon hides information, we could associate to it an "entanglement" entropy and, further, energy [8].

The origin of entropy is in the vacuum fluctuations of quantum fields, fluctuations which acquire a thermal character from the point of view of a uniformly accelerated (Rindler) observer, with the temperature [9]

$$T = g/2\pi, \qquad (1)$$

where "g" is the rest-system acceleration, i.e. the modulus of the acceleration 4 – vector measured by an inertial observer instantaneously at rest with the accelerated one. As Padmanabhan has noticed, we might associate an entropy to any surface, even in flat space, because is always possible to find a Rindler frame such that the chosen surface acts as a horizon for some Rindler observer. Therefore, any surface in Minkowski space must have an entropy. But for an accelerated observer the distance to the horizon is 1/g in cartesian coordinates, so that

$$g^{-1} = \Delta t. \qquad (2)$$

Putting (2) in the Davies – Unruh formula (1), we get, for "the energy per particle"

$$\varepsilon = (3/2)T = 3/4\pi\Delta t.$$

It yields

$$\varepsilon \Delta t \approx 1, \qquad (3)$$

a Heisenberg type relation. It means the shorter the time of measurement, the larger the energy fluctuations.

We chose to study $\Lambda(t)$ in (static) spherical Rindler coordinates [10], to avoid the special direction given by acceleration

$$ds^2 = -g^2 \xi^2 \cos^2\theta \, dt^2 + d\xi^2 + \xi^2 \, d\Omega^2. \qquad (4)$$

It is a well known result that an accelerated observer detects particles as if it were in a thermal heat bath at the temperature $g/2\pi$ [9]. Why an Unruh detector click when it is accelerated, even in Minkowski space ? . Basing on the fact that v.e.v. $<T_{\mu\nu}>$ is a covariant object, the answer would be "no". If the regularized stress tensor $<T_{\mu\nu}>$ vanishes in one frame (say the Minkowski frame) then it must vanish in all frames and hence in the Rindler frame. On the other hand, the detector would not have to click as long as it will never see any curvature of spacetime.

We propose to avoid the contradiction by replacing the flat space with a conformally flat one, the conformal factor being Lorentz – invariant.

Let us consider the Lorentzian version of the Hawking wormhole [11], which is Minkowski space far from the light cone

$$ds^2 = \left(1 - \frac{b^2}{x_\alpha x^\alpha}\right)^2 \eta_{\mu\nu}\, dx^\mu dx^\nu, \tag{5}$$

where "b" is the neck's radius of the wormhole (which will be taken of the order of the Planck length), $\acute{\eta}_{\mu\nu}$ = diag(-1, 1, 1, 1) and $x^{\acute{\alpha}} x_{\acute{\alpha}}$ ($\acute{\alpha}$ = 0,1,2,3) is the square of the Minkowski interval. The metric (5) is a solution of the Einstein's equations coupled to a conformally massless scalar field [12]. In spherical Rindler coordinates, eq.(5) becomes

$$ds^2 = \left(1 - \frac{b^2}{\xi^2}\right)^2 \left(-g^2 \xi^2 \cos^2\theta\, dt^2 + d\xi^2 + \xi^2 d\Omega^2\right). \tag{6}$$

We have horizons at $\xi$ = b and $\theta = \pi/2$ (because of the conformal factor, $\xi$ = 0 is no longer a horizon as in the Rindler geometry).

Only the region $\xi \geq b$ will be studied in this letter.

On the grounds of the importance of the surface integral in expressing the ADM energy or the entropy of a black hole, it seems suitable to take $\Lambda$ to be the Gaussian curvature of the causal horizon of the spacetime. Since the observer appears to be inside a box with dimension $\Delta t$, we put
$$\Lambda = (\Delta t)^{-2}. \tag{7}$$

With, for instance, $\Delta t \sim 10^{-12}$ s, the order of an atomic transition, the radius of the instantaneous Universe is 0.3 mm while the corresponding $\Lambda \sim 10^3$ cm$^{-2}$. For the electroweak scale, $\Delta t \sim 10^{-26}$ s, which leads to $\Lambda \sim 10^{32}$ cm$^{-2}$.

Let us now compute $\Lambda(t)$ for the spacetime (6). Thanks to the Landau and Lifshitz prescription [13], we have for the Gaussian curvature

$$k = (1/\sigma) R_{\theta\varphi\theta\varphi} = \xi^{-2}\left(1 - b^2/\xi^2\right)^{-2}, \tag{8}$$

where $\xi = \xi(t)$, $\sigma$ is the determinant of the two – surface $\xi$ = const., t = const. and $R_{\theta\varphi\theta\varphi}$ is the corresponding component of the Riemann tensor.

The next step is to calculate $\xi(t)$ from (8). Keeping în mind that the causal horizon means the two – surface on which the photon travels, what we have to do is to compute the null radial geodesic

---

[1] We could consider $\theta = \theta_0$, but things do not change too much (it means to replace g by $g\cos\theta_0$).

în the geometry (6).

One geodesic is, of course, the event horizon $\xi = b$. Taking then $\theta = 0$[1] and $\varphi = $ const., we obtain

$$g^2 \xi^2 dt^2 + d\xi^2 = 0,$$

whence

$$\xi^{\pm}(t) = \xi_0 e^{\pm gt}. \tag{9}$$

A comparison with the Minkowski coordinates leads to[2] $\xi_0 = 1/g$ [10]. The two signs în eq. (9) correspond to the past and future null rays.

We have for $\Lambda^+(t)$ the expression

$$\Lambda^+(t) = \frac{g^2 e^{-2gt}}{(1-b^2 g^2 e^{-2gt})^2}. \tag{10}$$

From $\xi \geq b$ we obtain exp $(gt) > bg$ and $\Lambda^+$ is always finite (excepting on the event horizon $\xi = b$). It reaches the maximum value $g^2(1-b^2g^2)^{-2}$ at $t = 0$[3] and is vanishing at infinity. For example, after $t_1 = 10^{-10}$ s and with $g_1 = 3.10^{20}$ cm$^{-2}$, we have $gt = 1$ but $\Lambda^+ \sim 0.01$ cm$^{-2}$. After $t_2 = 1$s, $\Lambda^+$ is already much smaller.

As far as the associated energy $W(t)$ is concerned, we have

$$W^+(t) = \frac{g e^{-gt}}{1 - b^2 g^2 e^{-2gt}}. \tag{11}$$

The fact that $W$ is proportional to $\Lambda^{1/2}$ is a consequence of (7) combined with the Heisenberg Principle. $W^+(t)$ decreases from the initial value $g/(1-b^2g^2)$ to zero at infinity.

For $g \ll 1/b$, $W^+(t)$ is independent of the Newton constant $G$. It depends only on $\hbar$ and $c$. For example, with the previous values $g_1$ and $t_1$ we get $W^+_1 = 10^{-17}$ ergs, on the causal horizon $\xi_1 = 8$ cm.

For the past null rays, we have $\xi = \xi^-$ and

$$\Lambda^-(t) = \frac{g^2 e^{2gt}}{\left(1 - b^2 g^2 e^{2gt}\right)^2}. \tag{12}$$

---

[2] Taking into consideration that g is the proper acceleration of an observer sitting at $x^3 = 0$ in Cartesian coordinates (namely, $\xi = 1/g$), the choice $\xi_0 = 1/g$ is most suitable.

[3] We are dealing with accelerations smaller then the Planck value $1/b$.

We have now exp (gt) < 1/bg, from $\xi^- > b$. Therefore, $0 < t < (-1/g) \ln bg$. In this case $\Lambda^-$ is increasing from $g^2(1-b^2g^2)^{-2}$ at t = 0, to infinity at $t_{max} = (-1/g)\ln bg$ or $\xi^- = b$, where the Gauss curvature is also infinite ($\xi^-$ is decreasing from 1/g to the Planck length b).

A similar dependence with respect to time we have for $W^-(t)$

$$W^-(t) = \frac{g\, e^{gt}}{1-b^2 g^2 e^{2gt}}, \tag{13}$$

with $W^-(0) = g/(1-b^2g^2)$ and $W^-(t_{max})$ – infinite.

We note that, on the second hemisphere ($\pi/2 \leq \theta \leq \pi$), $\cos\theta_0$ changes its sign. This is equivalent to a formal sign change of g. Therefore, in this region, $\Lambda^+$ becomes $\Lambda^-$.

It is instructive to calculate "the cosmological constant" for the de Sitter (dS) metric[4]

$$ds^2 = -dt^2 + e^{2Ht}(dr^2 + r^2 d\Omega^2), \tag{14}$$

where $H = 8\pi G\rho/3 = $ const. Using eq.(8) we obtain, for the Gaussian curvature

$$k = r^{-2}(t)\, e^{2Ht}, \tag{15}$$

r(t) being the null radial geodesic. We have, from eq. (14)

$$r(t) = H^{-1}(1 - e^{-Ht}) \tag{16a}$$

for dr/dt > 0 (outgoing null rays) and

$$r(t) = H^{-1} e^{-Ht} \tag{16b}$$

for dr/dt < 0 (ingoing null rays).
Eq.(15) yields

$$\Lambda^+(t) = H^2 (e^{Ht} - 1)^{-2} \tag{17a}$$

and

$$\Lambda^-(t) = H^2. \tag{17b}$$

We see that $\Lambda^+$ approaches the Minkowski value $t^{-2}$ for Ht << 1, when exp Ht ~ 1+Ht.
A similar dependence was obtained by O.Bertolami [14] in a Brans – Dicke theory with a scalar field $\phi(t)$ – dependent cosmological term, with "t" the cosmic time.

As far as the energy is concerned, $W^- = H^5$, while $W^+$ is given by

$$W^+(t) = H (e^{Ht} - 1)^{-1}. \tag{18}$$

---

[4] We consider the dS geometry was obtained from Einstein's equations with $p + \rho = 0$ as the equation of state of a perfect fluid as source.

When Ht << 1, we get $W^+(t) = 1/t$, as per the Minkowski case and in accordance with the Heisenberg Principle.

To summarize, we stress that our $\Lambda(t)$ is neither cosmological, nor constant. It depends on the spacetime we are living in, being nonvanishing even in Minkowski space. Its origin comes from the vacuum fluctuations of quantum fields. The corresponding energy is proportional to $\Lambda^{1/2}$ and is localized on the null horizon.

[5]The fact that $W^-$ is constant is in accordance with the interpretation of $\Lambda$ as due to vacuum fluctuations of quantum fields. The dS space is expanding and, because of its horizon, the Casimir effect leads to a decreasing of $W^-$. But the null rays are converging to the origin and the two effects are compensated for.